\documentclass[a4paper]{jpconf}
\usepackage{graphicx}
\usepackage[percent]{overpic}

\begin{document}
\title{Heavy flavor production in the STAR experiment}

\author{Barbara Trzeciak$^1$ for the STAR Collaboration}

\address{$^1$Faculty of Nuclear Sciences and Physical Engineering, Czech Technical University in Prague, Brehova 7, 115 19 Praha 1, Czech Republic}

\ead{trzecbar@fjfi.cvut.cz}

\begin{abstract}
In this paper, recent STAR heavy flavor measurements in proton-proton and heavy-ion collisions are highlighted. We report studies of open charm mesons, reconstructed directly from hadronic decay products, and studies of electrons from semi-leptonic decays of heavy flavor hadrons. We also present J/$\psi$ measurements via the di-electron decay channel at various collision systems and energies. In Au+Au collisions the energy dependence of J/$\psi$ production measured at $\sqrt{s_{NN}}$ = 39, 62.4 and 200 GeV is shown. Finally, prospects of heavy flavor measurements with the STAR detector upgrades are discussed.
\end{abstract}

\section{Introduction}
Heavy quarks are produced at the initial hard-scattering stage of the ultra-relativistic heavy-ion collisions at Relativistic Heavy Ion Collider (RHIC). Heavy quarks are expected to interact with the hot and dense medium created in RHIC high energy collisions differently from light quarks therefore they serve as unique probes of the properties of the Quark-Gluon Plasma (QGP). Heavy quarks may also help to understand the parton energy loss mechanism and the degree of the medium thermalization since their production and elliptic flow are sensitive to the dynamics of the medium. 

Moreover, studies of production of various quarkonium states in heavy-ion collisions can provide insight into the thermodynamic properties of the hot and dense medium. It was proposed that due to a Debye screening of the quark-antiquark potential in the hot medium quarkonia are dissociated and thus this ''melting'' can be a signature of the QGP formation. Also, since different quarkonium states have different binding energies they are expected to dissociate at different temperatures and can be treated as a QGP thermometer. But there are other mechanisms that can alter quarkonium yields in heavy-ion collisions relative to $p+p$ collisions, for example cold nuclear matter effects (shadowing/anti-shadowing, final state nuclear absorption) or statistical coalescence of heavy quark-antiquark pairs.
Systematic measurements of the quarkonium production for different colliding systems, centralities and collision energies may help to understand the quarkonium production mechanisms in heavy-ion collisions as well as the medium properties.

\section{Open Heavy Flavor}

\subsection{D mesons}

Open charm mesons can be directly reconstructed via their hadronic decays products (e.g. $D^{0} \rightarrow K^{-} + \pi ^{+} $). In this way, a direct access to the heavy flavor hadron kinematics is obtained. But it is difficult to trigger and the reconstruction is challenging without a good secondary vertex resolution. Since STAR did not have a good secondary-vertex detector when the data that are presented in this paper were recorded, the large combinatorial background contribution is estimated and subtracted on statistical basis using particle combinations with wrong charge signs or from different events.

\begin{figure}[htbp]
	\begin{minipage}[b]{0.49\linewidth}
		\centering
		\includegraphics[width=1.\textwidth]{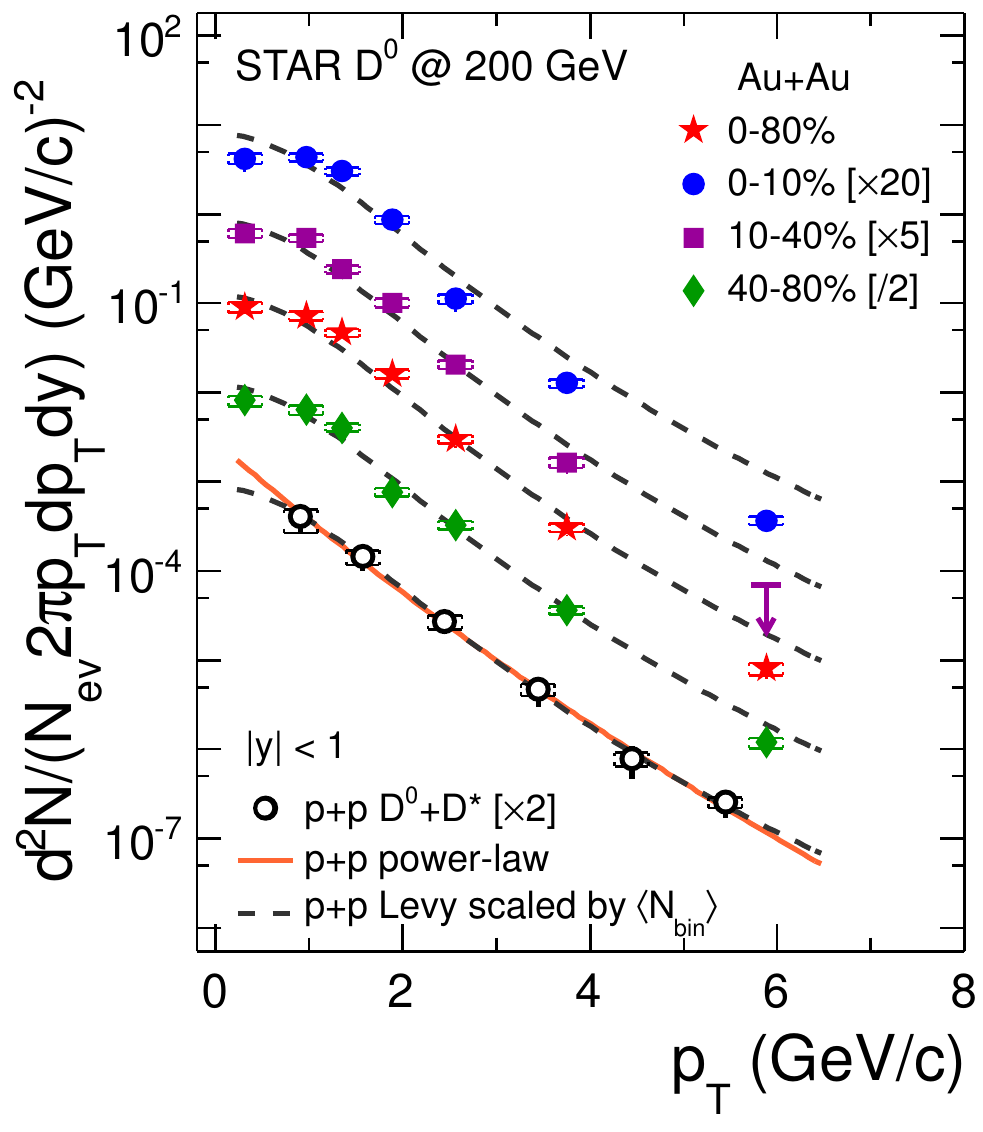}
		\caption{$D^{0}$ $p_{T}$ spectra in Au+Au collisions (full symbols) and $p+p$ collisions (open symbols) at $\sqrt{s_{NN}} =$ 200 GeV at mid-rapidity in different centrality bins \cite{Adamczyk:2014uip}. Dashed lines represent Levy function fit to $p+p$ scaled by the number of binary collisions.}
		\label{fig:D0_pt}
	\end{minipage}
	\hspace{.2in} 
	\begin{minipage}[b]{0.49\linewidth}
		\centering
		\includegraphics[width=0.95\linewidth]{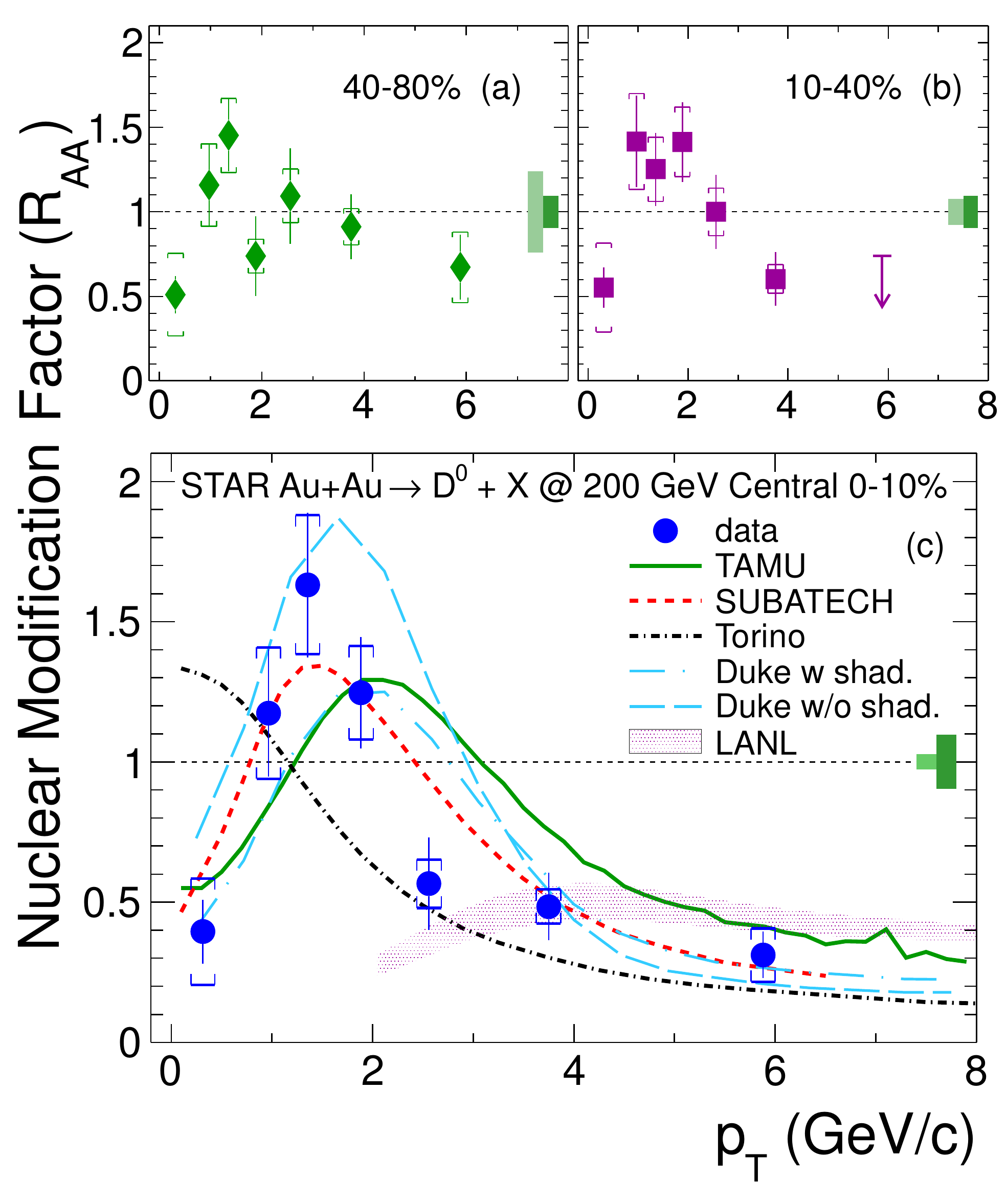}
		\caption{$D^{0}$ $R_{AA}$ as a function of $p_{T}$ for Au+Au collisions at $\sqrt{s_{NN}} =$ 200 GeV at mid-rapidity with different model calculations \cite{Adamczyk:2014uip}. \\ \\}
		\label{fig:D0_Raa}
	\end{minipage}
\end{figure}

Figure \ref{fig:D0_pt} presents the $D^{0}$ $p_{T}$ spectra in Au+Au collisions at $\sqrt{s_{NN}} =$ 200 GeV at mid-rapidity for different centralities \cite{Adamczyk:2014uip}. The $D^{0} + D^{*}$ measurement in p+p collisions at $\sqrt{s} =$ 200 GeV \cite{Adamczyk:2012af} is also shown as open circles. The dashed curves represent Levy functions fitted to the $p+p$ data and scaled up by the number of binary collisions ($N_{bin}$) for each centrality bin. The charm production cross-section at mid-rapidity in the 0-10\% central collisions is found to be $d\sigma_{c \bar{c} } ^ {NN}/dy = $ 148 $\pm$ 15 (stat.) $\pm$ 19 (sys.) $\mu b$ and the cross-sections per nucleon-nucleon collision show a number-of-binary-collision scaling from $p+p$ to Au+Au that indicates that charm quarks are mostly produced in the initial hard processes \cite{Adamczyk:2014uip}.

\begin{figure}[htbp]
		\centering
		\includegraphics[width=1.\textwidth]{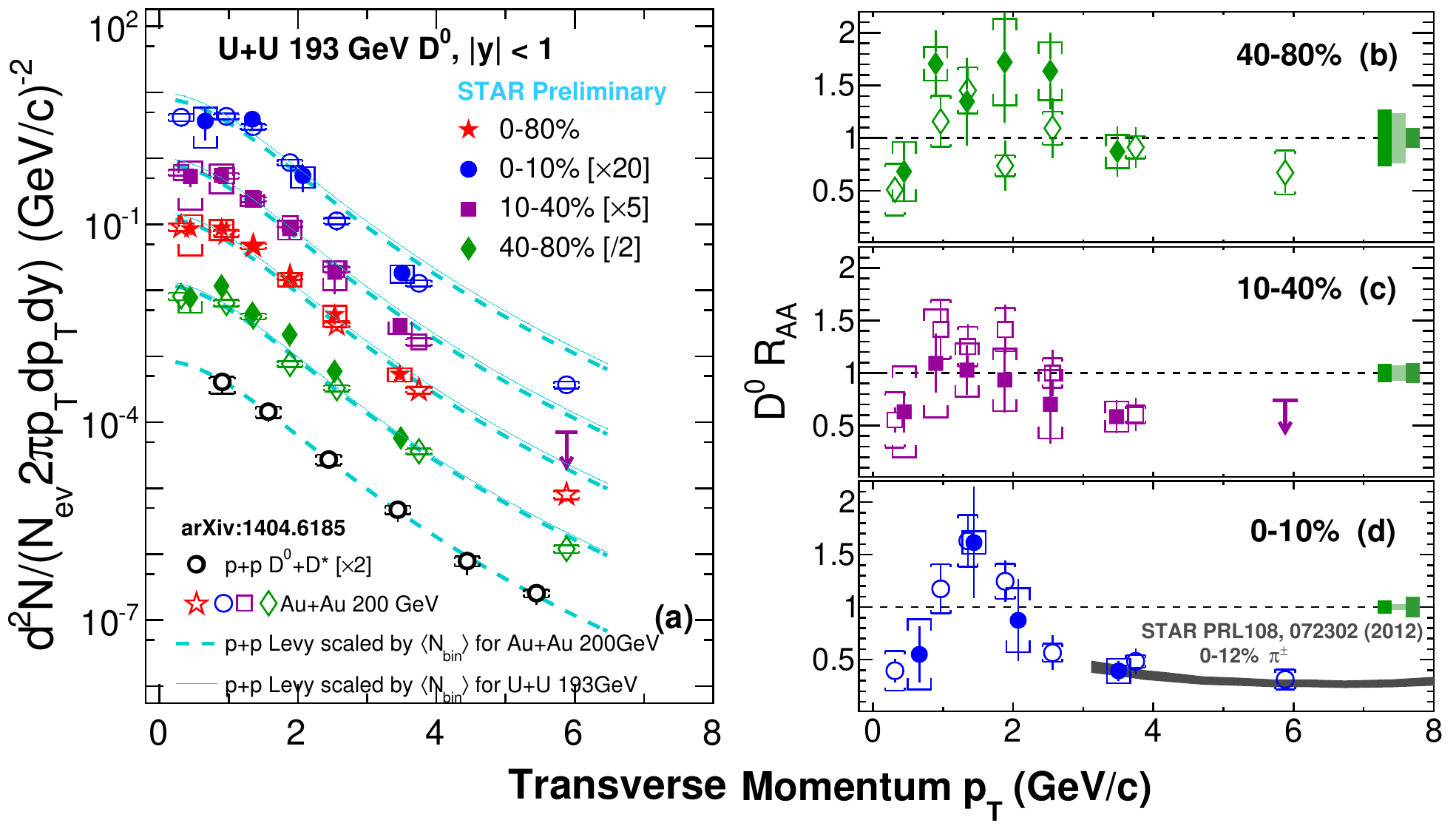}
		\caption{$D^{0}$ $p_{T}$ spectra (left panel) and $R_{AA}$ (right panel) in U+U collisions (full symbols) at $\sqrt{s_{NN}} =$ 193 GeV and Au+Au collisions (open symbols) at $\sqrt{s_{NN}} =$ 200 GeV at mid-rapidity.  Solid and dashed lines represent Levy function scaled by the number of binary collisions for U+U and Au+Au, respectively.}
		\label{fig:D0_UU}
\end{figure}

In the most central Au+Au collisions a strong suppression at high $p_{T}$ compared with the corresponding Levy function is observed. To quantify the suppression, the nuclear modification factor ($R_{AA}$) is calculated and shown in Fig. \ref{fig:D0_Raa} as a function of $p_{T}$ for different centrality bins \cite{Adamczyk:2014uip}. No suppression is seen for peripheral collisions, while for the most central collisions the $R_{AA}$ is about 0.5 for $p_{T} >$ 3 GeV/$c$. This suppression at high $p_{T}$ is also consistent with those of non-photonic electrons (see Sec. \ref{sec:nonPhotonic}, shown in Fig. \ref{fig:NPA_Raa}) and light hadrons, $\pi^{\pm}$, (shown in Fig. \ref{fig:D0_UU}(d)). The 0-10\% result is compared with different model predictions \cite{He:2011qa, He:2012df, Gossiaux:2010yx, Gossiaux:2012ya, Alberico:2011zy, Alberico:2013bza, Cao:2013ita, Sharma:2009hn}. The mid-$p_{T}$ enhancement seen in the data can be described by models that include strong charm-medium interactions and charm quark hadronization via coalescence with light quarks. 

Figure \ref{fig:D0_UU} shows $D^{0}$ $p_{T}$ spectra and $R_{AA}$ as a function of $p_{T}$ measured in U+U collisions at $\sqrt{s_{NN}} =$ 193 GeV at mid-rapidity (full symbols) together with the mid-rapidity Au+Au result (open symbols). In the same centrality bin, energy density in U+U collisions could be 20\% higher than in Au+Au collisions \cite{Kikola:2011zz}. Data show similar $R_{AA}$ trend in U+U and Au+Au collisions.

\subsection{Non-photonic electrons}\label{sec:nonPhotonic}

Open heavy flavor mesons can also be studied using electrons from their semi-leptonic decays (non-photonic electrons, NPE). The branching ratio for this process is higher and it is relatively easy to trigger on high-$p_{T}$ electrons. The disadvantage of this method is that it gives an indirect access to the heavy flavor hadron kinematics and it contains contributions from both charm and beauty hadron decays.

\begin{figure}[htbp]
	\begin{minipage}[b]{0.49\linewidth}
		\centering
		\includegraphics[width=1.\textwidth]{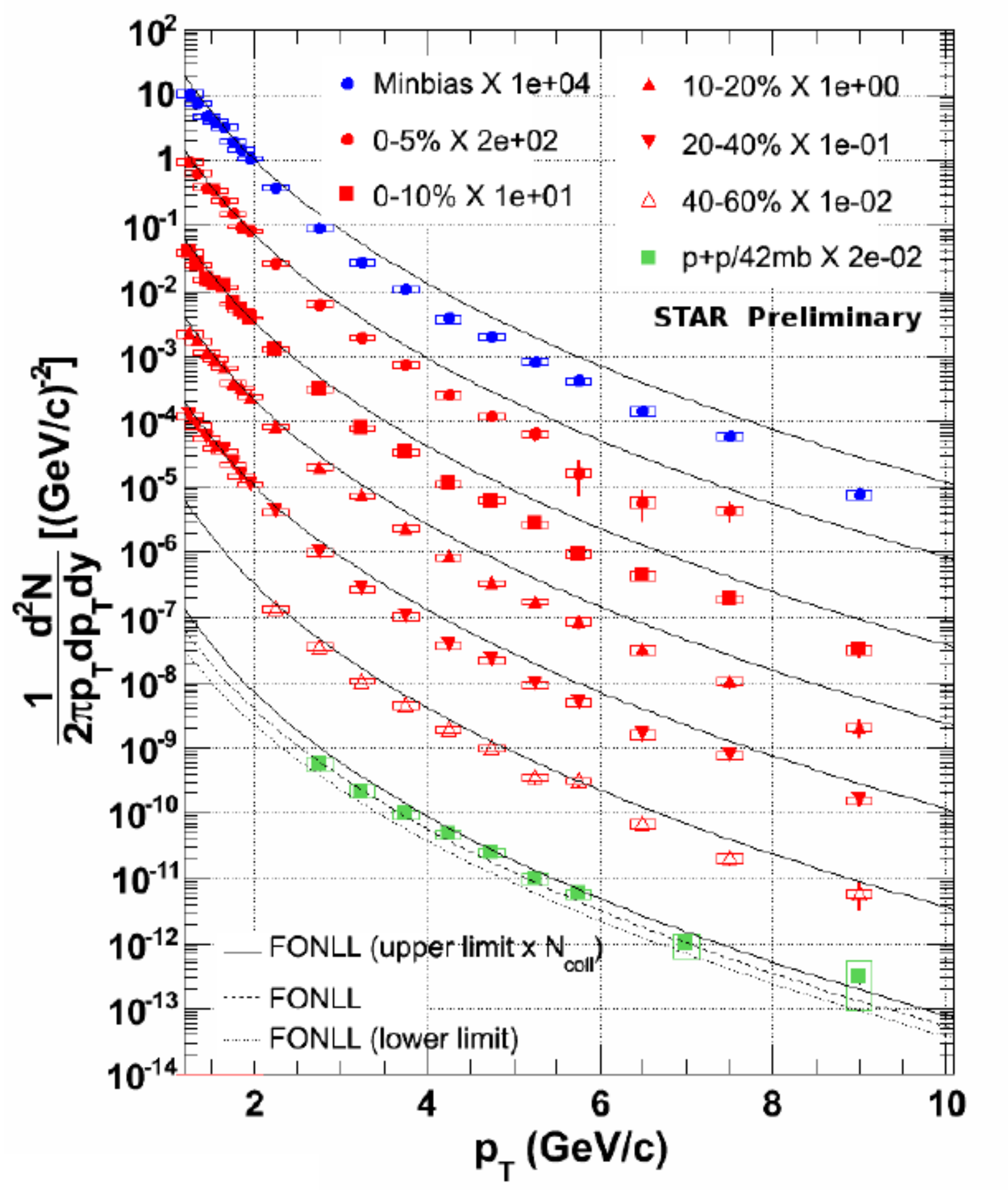}
		\caption{NPE $p_{T}$ spectra in Au+Au collisions for different centrality bins and $p+p$ collisions (green squares) at $\sqrt{s_{NN}} =$ 200 GeV at mid-rapidity. Solid lines represent the upper limit of FONLL calculations scaled by a number of binary collisions in a given centrality.}
		\label{fig:NPE_pt200}
	\end{minipage}
	\hspace{.2in} 
	\begin{minipage}[b]{0.49\linewidth}
		\centering
		\includegraphics[width=1.\linewidth]{NPE_pTSpectra62.pdf}
		\caption{NPE $p_{T}$ spectra in Au+Au collisions at $\sqrt{s_{NN}} =$ 62.4 GeV at mid-rapidity. Solid lines represent the upper limit of FONLL calculations scaled by a number of binary collisions in a given centrality. ISR results \cite{Basile:1981dn} are also shown.}
		\label{fig:NPE_pt62}
	\end{minipage}
\end{figure}

NPE $p_{T}$ spectra for Au+Au collisions at $\sqrt{s_{NN}} =$ 200 GeV at mid-rapidity together with the FONLL upper limit calculations \cite{Cacciari:2005rk} scaled up by $N_{bin}$ are shown in Fig. \ref{fig:NPE_pt200} for different centralities. The NPE production in the most central and mid-central collisions is suppressed compared to the FONLL calculations.
STAR has also obtained the first result of the NPE production in Au+Au collisions at $\sqrt{s_{NN}}$ = 62.4 GeV, shown in Fig. \ref{fig:NPE_pt62}. Solid and dashed lines represent the upper-limit and the central value of FONLL calculations scaled by $N_{bin}$, respectively. In contrast to the result in 200 GeV, no suppression at $p_{T} < $ 5.5 GeV/$c$ is observed compared to the FONLL calculations. The result is not corrected yet for the J/$\psi$ and Drell-Yan contributions. The J/$\psi$ contribution is about 20\% at high $p_{T}$ at $\sqrt{s_{NN}} =$ 200 GeV. Even taking this into account, no significant suppression would be seen at 62.4 GeV.

\begin{figure}[htbp]
	\begin{minipage}[b]{0.49\linewidth}
		\centering
		\begin{overpic}[width=1.\textwidth]{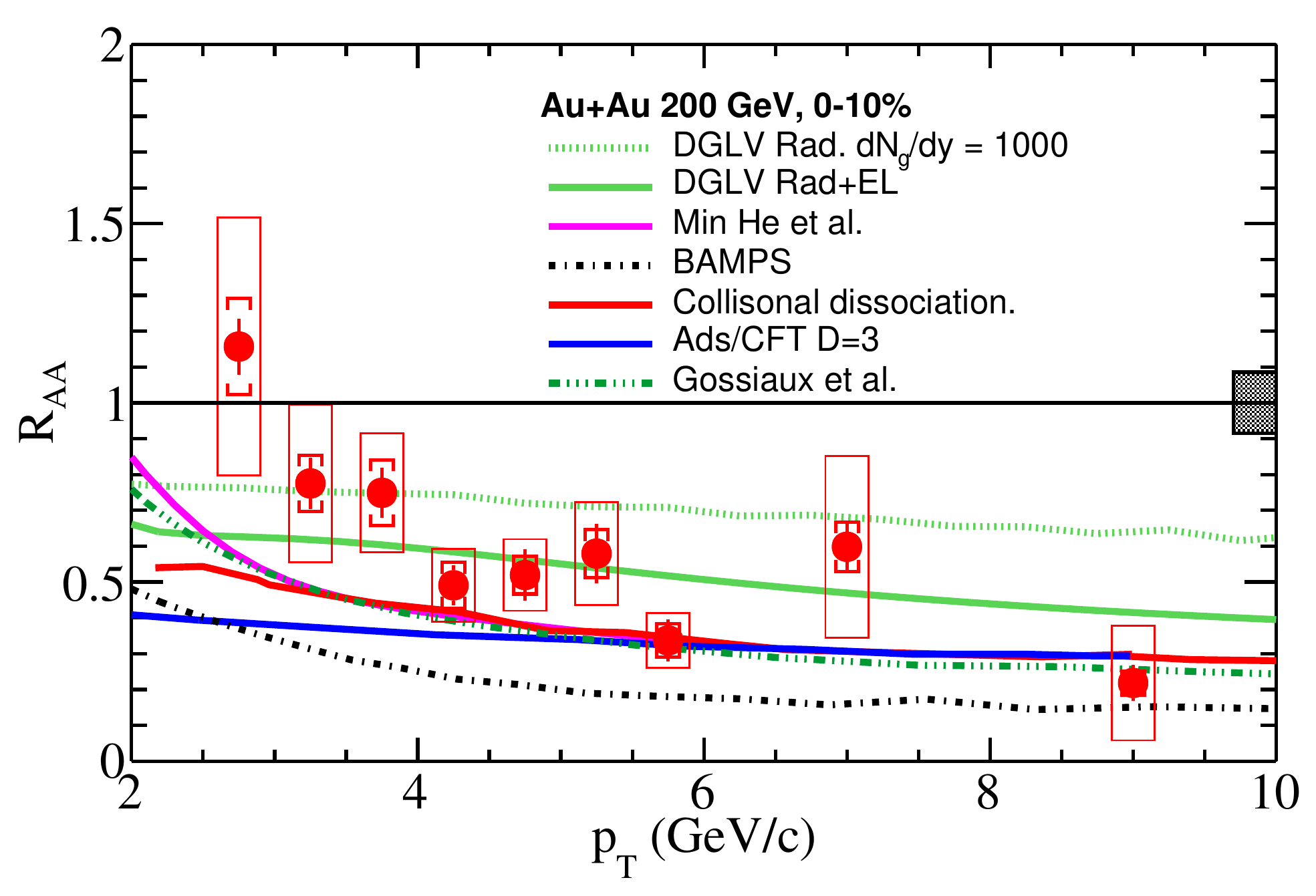}
		\put(15,60) {\footnotesize$STAR$}
		\put(15,55) {\footnotesize$Preliminary$}
		\end{overpic}
		\caption{NPE $R_{AA}$ as a function of $p_{T}$ for Au+Au collisions at $\sqrt{s_{NN}} =$ 200 GeV at mid-rapidity in 0-10\% most central events with different model predictions. }		
		\label{fig:NPA_Raa}		
	\end{minipage}
	\hspace{.2in} 
	\begin{minipage}[b]{0.49\linewidth}
		\centering
		\includegraphics[width=0.95\linewidth]{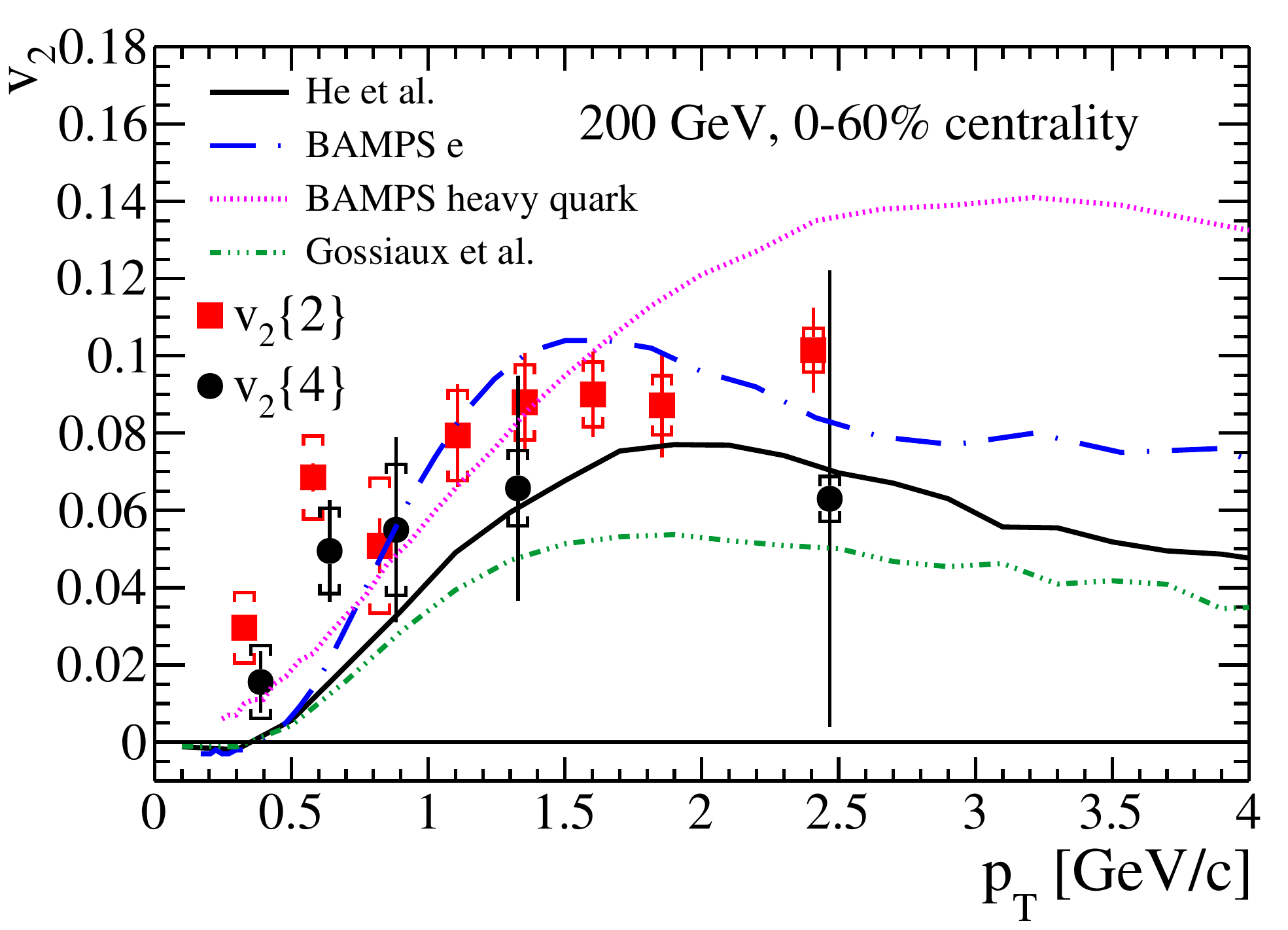}
		\caption{NPE $v_{2}$ for Au+Au collisions at $\sqrt{s_{NN}} =$ 200 GeV at mid-rapidity in 0-60\% central events with different model predictions \cite{Adamczyk:2014yew}.}
		\label{fig:NPE_v2}
	\end{minipage}
\end{figure}

NPE $R_{AA}$ in the most central Au$+$Au collisions (0-10\%) at $\sqrt{s_{NN}}$ = 200 GeV at mid-rapidity is shown in  Fig. \ref{fig:NPA_Raa} as a function of $p_{T}$. A strong suppression, similar to above-mentioned $D^{0}$ result, is seen at high $p_{T}$. Figure \ref{fig:NPE_v2} shows the measured elliptic flow of NPE in Au+Au collisions at $\sqrt{s_{NN}}$ = 200 GeV at mid-rapidity \cite{Adamczyk:2014yew}. The measurement is done using the two- ($v_{2}\lbrace 2 \rbrace$) and four-particle ($v_{2}\lbrace 4 \rbrace$) correlations. $v_{2}\lbrace 2 \rbrace$ and $v_{2}\lbrace 4 \rbrace$ are finite at $p_{T} >$ 0.5 GeV/$c$ and results obtained from both methods are consistent with each other. Also, at $p_{T} >$ 4 GeV/$c$ an increase of $v_{2}$ is observed that is probably due to jet-like correlations \cite{Adamczyk:2014yew}. 

The measured $R_{AA}$ and $v_{2}$ are compared with different model predictions. The gluon radiation scenario alone (DGLV) \cite{Djordjevic:2005db} fails to describe the large NPE suppression while adding collisional energy loss improves the agreement with the data. The $R_{AA}$ can be described by the collisional dissociation model \cite{Sharma:2009hn} and AdS/CFT calculations \cite{Horowitz:2007ui}.
The partonic transport model, BAMPS \cite{Uphoff:2011ad, Uphoff:2012gb} describes the $v_{2}$ data well but it underpredicts the NPE $R_{AA}$ result. The He {\it{et al.}} \cite{He:2011qa, vanHees:2007me} and the Gossiaux {\it{et al.}} \cite{Gossiaux:2010yx, Gossiaux:2008jv, Aichelin:2012ww} models underpredict $v_{2}\lbrace 2 \rbrace$ but they are in a good agreement with the measured NPE $R_{AA}$. The Gossiaux {\it{et al.}} model includes radiative and collisional energy loss and the He {\it{et al.}} model is a TMatrix interactions model. In general, it is challenging to describe well NPE $R_{AA}$ and $v_{2}$ simultaneously.

\section{Quarkonia}

Quarkonia at STAR are measured via the di-electron decay channel at mid-rapidity ($\vert y \vert < $ 1). 
STAR has observed significant suppressions of $\Upsilon$ production in central Au+Au collisions at $\sqrt{s_{NN}} =$ 200 GeV compared to $p+p$ collisions, more details can be found here \cite{Adamczyk:2013poh}.
Here, we will focus on low and high-$p_{T}$ inclusive J/$\psi$ measurements at different collision energies.

\begin{figure}[htbp]
	\begin{minipage}[b]{0.49\linewidth}
		\centering
		\includegraphics[width=1.\textwidth]{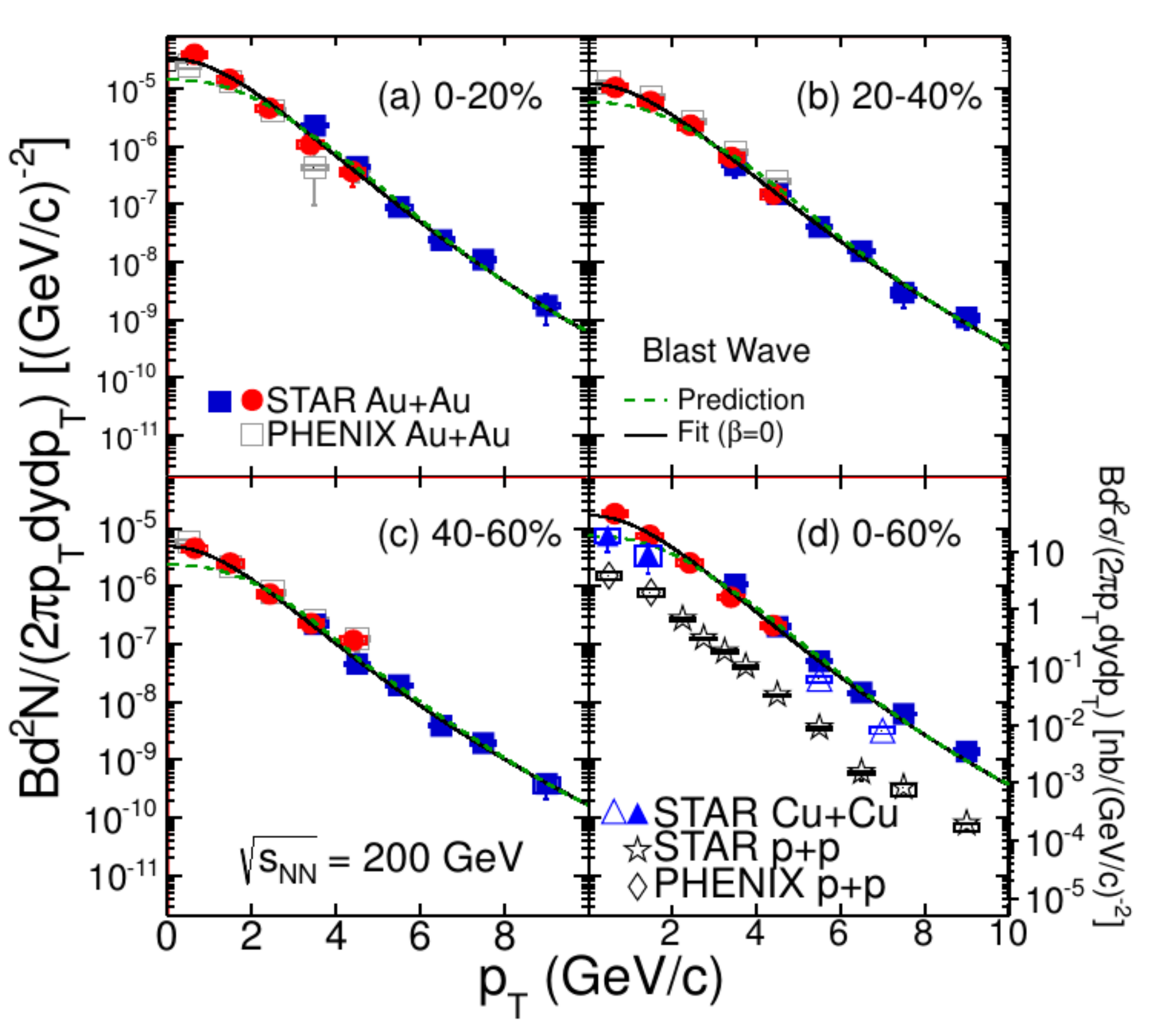}
		\caption{J/$\psi$ $p_{T}$ spectra in Au+Au collisions at $\sqrt{s_{NN}} =$ 200 GeV at mid-rapidity, for different centrality bins, with Tsallis Blast-Wave predictions. On panel (d) $p+p$ results as stars and Cu+Cu results as triangles are also presented.}
		\label{fig:Jpsi_pT}
	\end{minipage}
	\hspace{.2in} 
	\begin{minipage}[b]{0.49\linewidth}
		\centering
		\includegraphics[width=1.\linewidth]{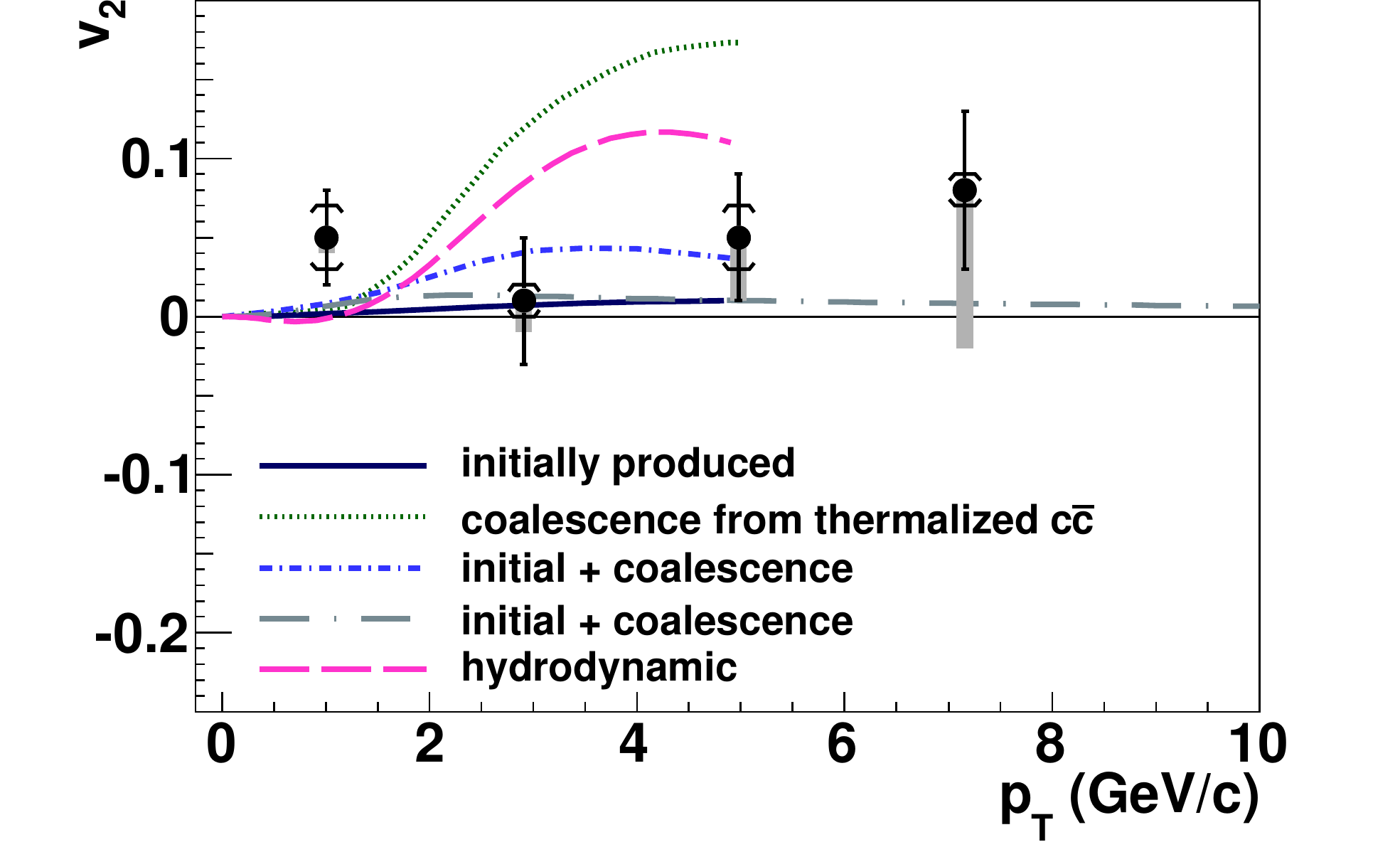}
		\caption{J/$\psi$ $v_{2}$ in Au+Au collisions at $\sqrt{s_{NN}}$ = 200 GeV at mid-rapidity in 0-80\% central events with different model predictions \cite{Adamczyk:2012pw}. \\ \\}
		\label{fig:Jpsi_v2}
	\end{minipage}
\end{figure}

Figure \ref{fig:Jpsi_pT} shows STAR J/$\psi$ $p_{T}$ spectra in Au+Au collisions at $\sqrt{s_{NN}} =$ 200 GeV for different centrality bins \cite{Adamczyk:2013tvk,Adamczyk:2012ey} as full symbols. The spectra are compared to the Tsallis Blast-Wave (TBW) prediction assuming that J/$\psi$ flows like lighter hadrons \cite{Tang:2008ud, Tang:2011xq}, shown as dashed lines. The J/$\psi$ $p_{T}$ spectra are softer than the TBW prediction 
$p_{T}$ and agree with TBW fit with radial flow $\beta =$ 0 (solid lines). This may indicate small J/$\psi$ radial flow or contribution from the recombination to low-$p_{T}$ J/$\psi$ production.

Measurement of J/$\psi$ $v_{2}$ may provide additional information about the J/$\psi$ production mechanisms. Figure \ref{fig:Jpsi_v2} shows J/$\psi$ $v_{2}$ measured in STAR in Au+Au collisions at $\sqrt{s_{NN}}$ = 200 GeV \cite{Adamczyk:2012pw}. At $p_{T} >$ 2 GeV/$c$ $v_{2}$ is consistent with zero. Comparing to different model predictions \cite{Yan:2006ve,Greco:2003vf,Zhao:2008vu,Liu:2009gx}, data disfavor the scenario that J/$\psi$ with $p_{T} >$ 2 GeV/$c$ are dominantly produced by coalescence from thermalized (anti-)charm quarks.

\begin{figure}[htbp]
	\begin{minipage}[b]{0.49\linewidth}
		\centering
		\includegraphics[width=0.85\textwidth]{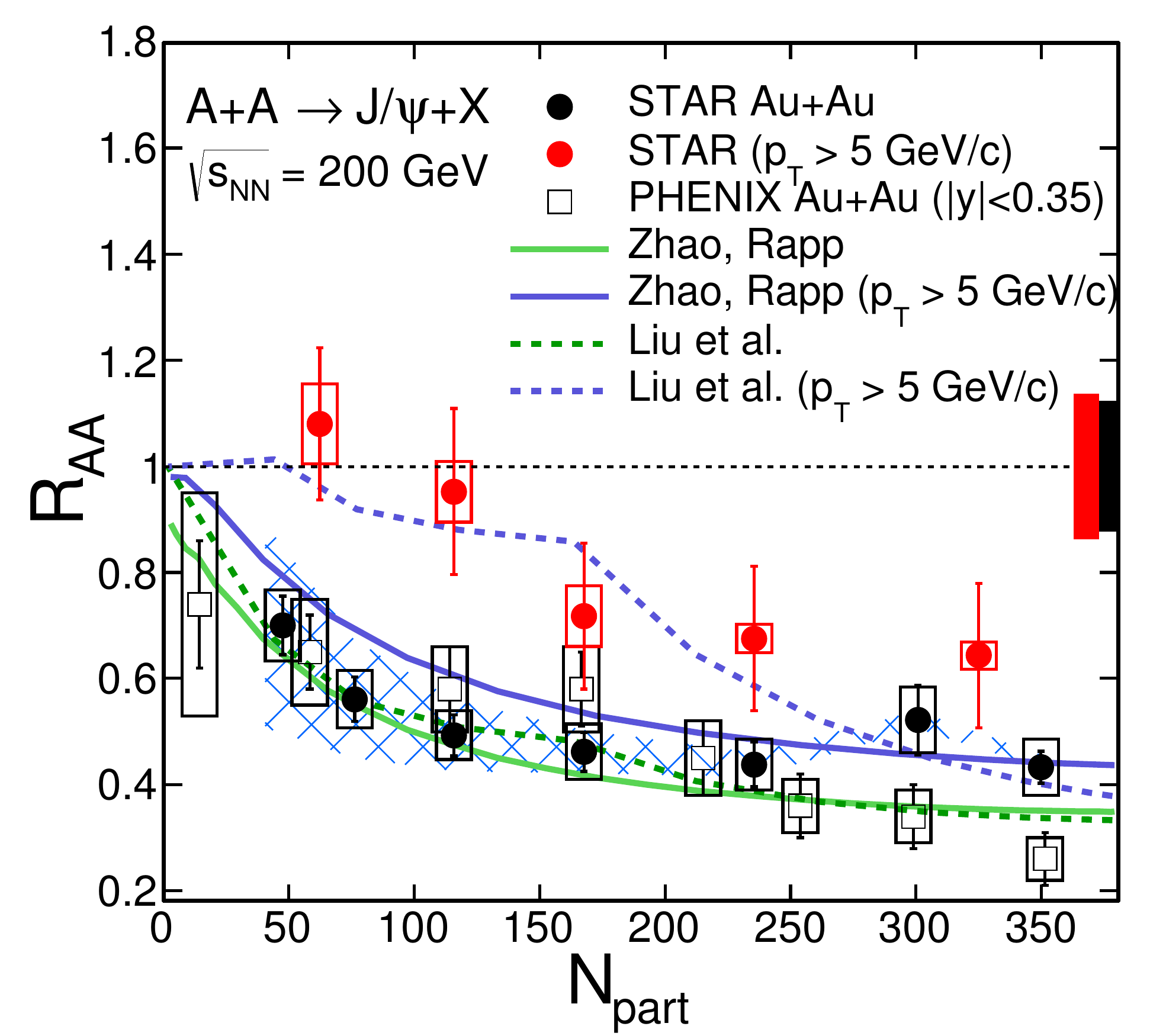}
		\caption{J/$\psi$ $R_{AA}$ as a function of $N_{part}$ in Au+Au collisions at $\sqrt{s_{NN}} =$ 200 GeV at mid-rapidity with two model predictions. The low-$p_{T}$ ($<$ 5 GeV/$c$) result is shown as black full circles and the high-$p_{T}$ ($>$ 5 GeV/$c$) measurement as red full circles. }
		\label{fig:Jpsi_Raa200}
	\end{minipage}
	\hspace{.2in} 
	\begin{minipage}[b]{0.49\linewidth}
		\centering
		\includegraphics[width=1.\linewidth]{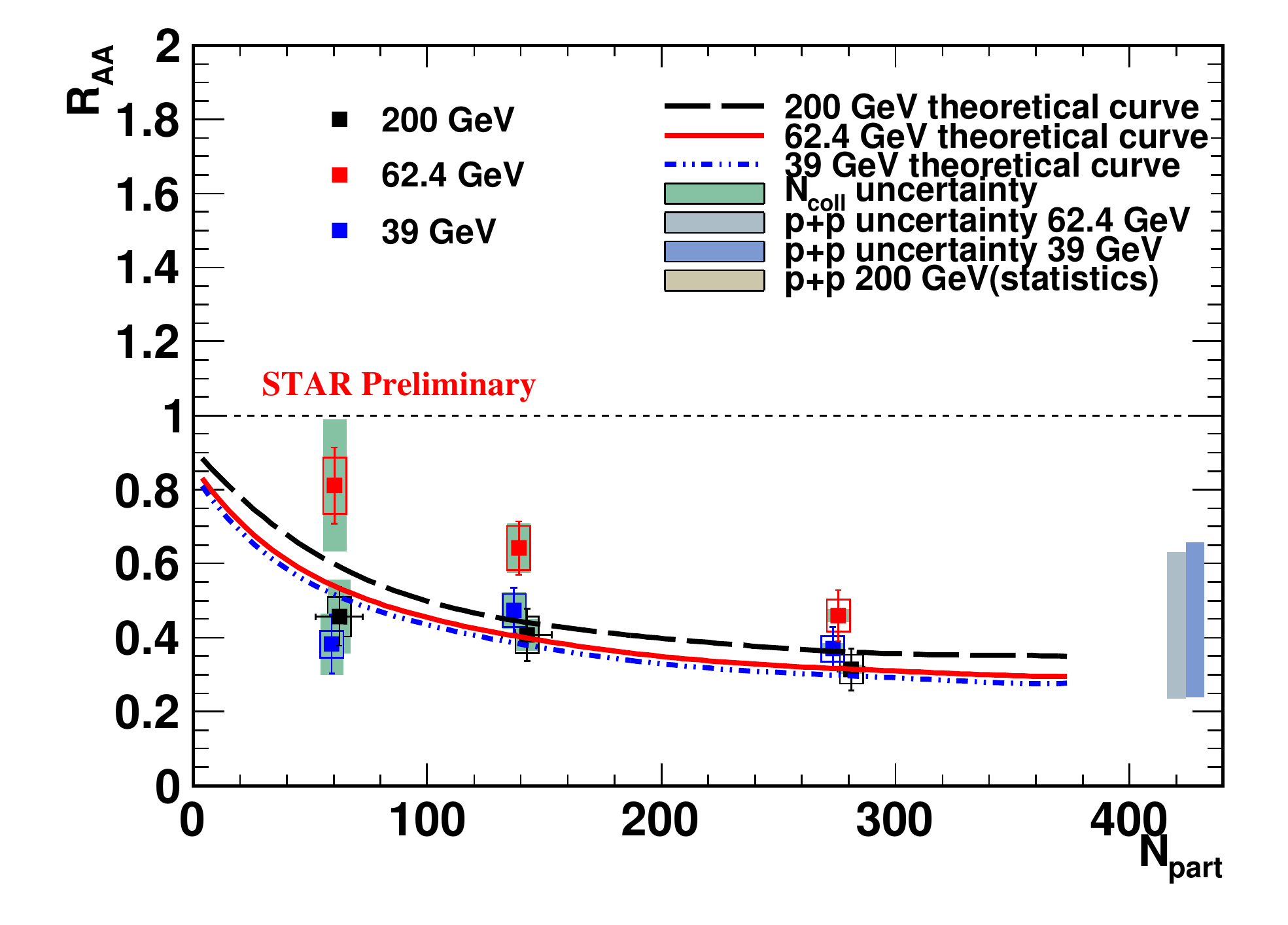}
		\caption{J/$\psi$ $R_{AA}$ as a function of $N_{part}$ in Au+Au collisions at $\sqrt{s_{NN}} =$ 200 (black), 62.4 (red) and 39 (blue) GeV at mid-rapidity with model predictions.\\ \\}
		\label{fig:Jpsi_RaaBES}
	\end{minipage}
\end{figure}

Figure \ref{fig:Jpsi_Raa200} shows J/$\psi$ $R_{AA}$ as a function of the number of participant nucleons ($N_{part}$) in Au+Au collisions at $\sqrt{s_{NN}}$ = 200 GeV. The low-$p_{T}$ ($<$ 5 GeV/$c$) result \cite{Adamczyk:2013tvk} is shown as black full circles and the high-$p_{T}$ ($>$ 5 GeV/$c$) measurement \cite{Adamczyk:2012ey} as red full circles. Suppression increases with collision centrality and $R_{AA}$ for high-$p_{T}$ J/$\psi$ is systematic higher than the low-$p_{T}$ J/$\psi$ $R_{AA}$. At high $p_{T}$ no significant suppression in peripheral and mid-central collisions is seen while in central collisions (0-30\%) a strong suppression is observed which may indicate color screening features - at $p_{T} >$ 5 GeV/$c$ the recombination and initial parton scattering effects are expected to be negligible \cite{Adamczyk:2012ey}. 

Figure \ref{fig:Jpsi_RaaBES} shows low-$p_{T}$ J/$\psi$ $R_{AA}$ in Au+Au collisions at lower collision energies, $\sqrt{s_{NN}}$ = 62.4 and 39 GeV together with the 200 GeV result. For lower initial energy densities the interplay between direct J/$\psi$ production and recombination from $c$ and $\bar{c}$ quarks is expected to be different. Similar level of suppression is seen in the data for all three energies, $\sqrt{s_{NN}}$ = 200, 62.4 and 39 GeV. However, one should note that due to lack of $p+p$ measurements at 62.4 and 39 GeV Color Evaporation Model calculations \cite{Nelson:2012bc} are used as baselines, which introduce large uncertainties shown as boxes in Fig. \ref{fig:Jpsi_RaaBES}.

The $R_{AA}$ results at $\sqrt{s_{NN}} =$ 200 GeV in Fig. \ref{fig:Jpsi_Raa200} are compared with two models, Zhao and Rapp \cite{Zhao:2010nk} and Liu {\it{et al.}} \cite{Liu:2009nb}. Both models take into account direct J/$\psi$ production with the color screening effect and J/$\psi$ produced via the recombination of $c$ and $\bar{c}$ quarks. Green lines represent low-$p_{T}$ predictions and both agree well with the measurement. High-$p_{T}$ predictions are shown as blue lines. At high $p_{T}$ the Liu {\it{et al.}} model describes the data well while the Zhao and Rapp model underpredicts the measured $R_{AA}$. For $\sqrt{s_{NN}}$ = 62.4 and 39 GeV the Zhao and Rapp model predicts similar suppression levels, and within the uncertainties the model agrees with the data (Fig. \ref{fig:Jpsi_RaaBES}).

\section{Outlook}

\begin{figure}[htbp]
	\begin{minipage}[b]{0.49\linewidth}
		\centering
		\includegraphics[width=1.\textwidth]{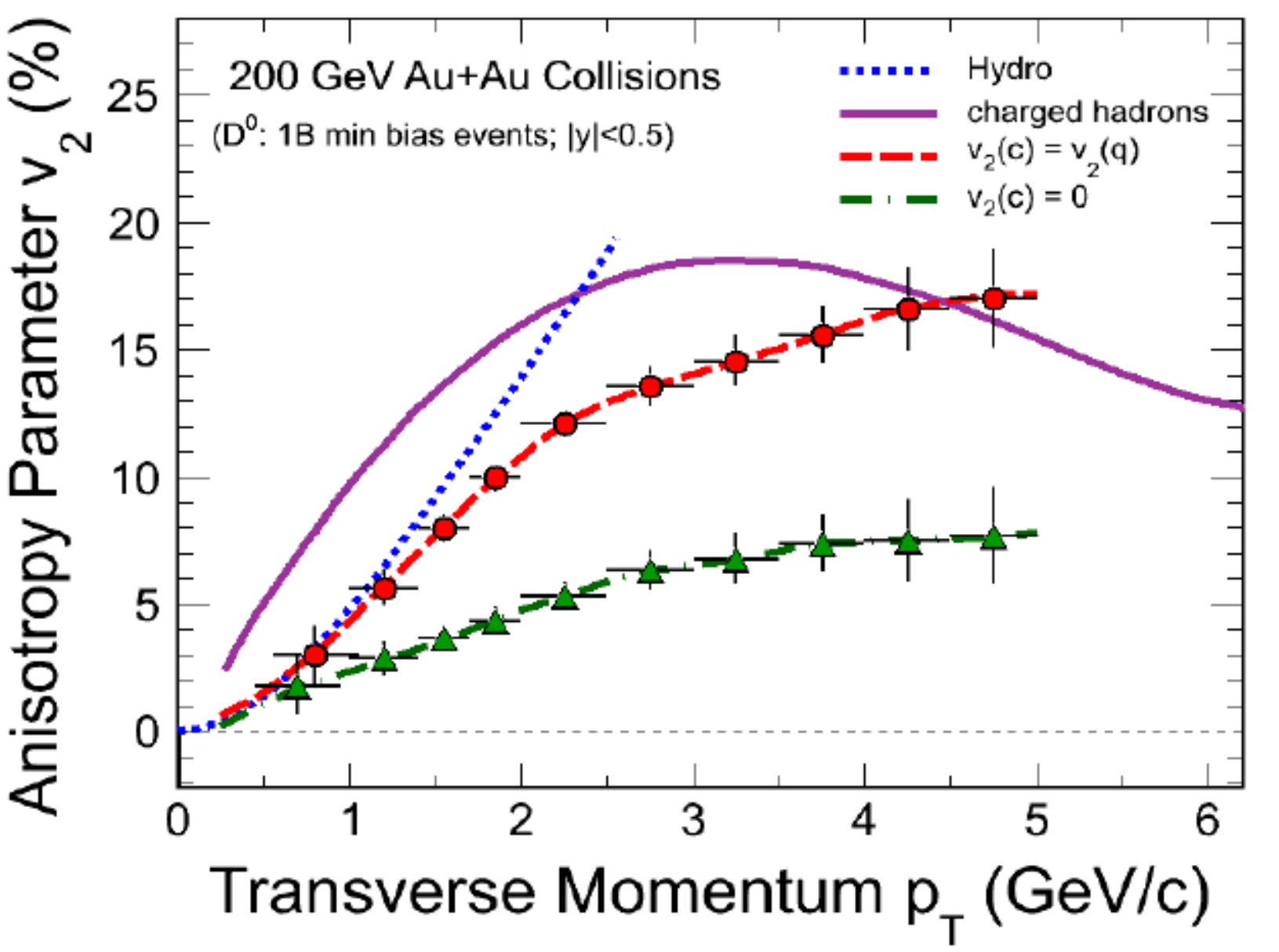}
		\caption{Projected $D^{0}$ $v_{2}$ measurement with the HFT detector, dashed lines represent different model assumptions. \\ \\ \\}
		\label{fig:D0_v2HFT}
	\end{minipage}
	\hspace{.2in} 
	\begin{minipage}[b]{0.49\linewidth}
		\centering
		\includegraphics[width=1.\linewidth]{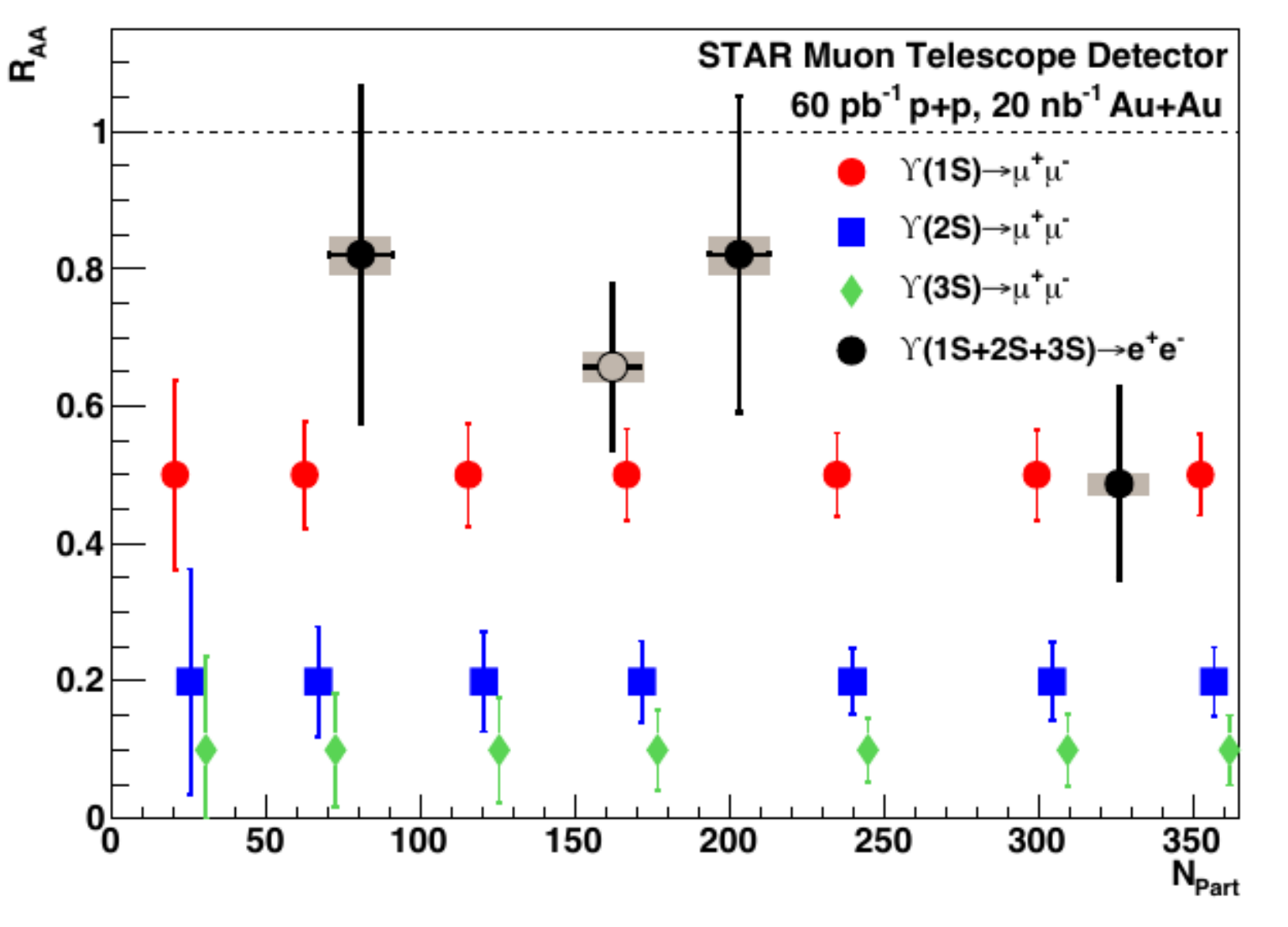}
		\caption{Statistical projection of $\Upsilon$ $R_{AA}$ measurement for different $\Upsilon$ states with the MTD detector. The figure also shows in black measured STAR $\Upsilon$ $R_{AA}$ \cite{Adamczyk:2013poh}, and has been modified from what was shown originally in order to include the published data.}
		\label{fig:UpsilonRaa_MTD}
	\end{minipage}
\end{figure}

STAR has two new detectors, fully installed and taking data in 2014: Heavy Flavor Tracker (HFT) and Muon Telescope Detector (MTD) \cite{Yaping:WWND2014}. The HFT is a silicon vertex detector. It allows precise open heavy flavor measurements, $v_{2}$ or $R_{AA}$, and to distinguish between charm and bottom. Projected $D^{0}$ elliptic flow $v_{2}$ measurement using the HFT detector is shown in Fig. \ref{fig:D0_v2HFT}. With the HFT it will also be possible to study non-prompt J/$\psi$ production from B meson decays using the displaced vertex. 
The MTD is situated outside the STAR magnet which is used as an absorber. The detector enables muon identification in STAR and so quarkonia measurements in the di-muon decay channel which is cleaner comparing with the di-electron decay channel. It has also excellent mass resolution that is particularly important for measurement of different $\Upsilon$ states. Figure \ref{fig:UpsilonRaa_MTD} shows precision projection for $R_{AA}$ of different $\Upsilon$ states with the MTD detector. With the MTD open heavy flavor studies can be also performed, using electron-muon correlations.

\section{Summary}

In summary, STAR observed strong NPE and $D^{0}$ suppression at high $p_{T}$ in Au+Au collisions at $\sqrt{s_{NN}}$ = 200 GeV, but no NPE suppression is seen at $\sqrt{s_{NN}}$ = 62.4 GeV at $p_{T} < $ 5.5 GeV/$c$ compare to pQCD calculations. Also, similar sizes of $D^{0}$ nuclear modification factors, $R_{AA}$, are seen in Au+Au and U+U collisions. $D^{0}$ $R_{AA}$ at $\sqrt{s_{NN}}$ = 200 GeV can be described by calculations that include strong charm-medium interactions and coalescence hadronization.
Significant suppression of low $p_{T}$ J/$\psi$ is seen at $\sqrt{s_{NN}}$ = 200, 62.4 and 39 GeV Au+Au collisions, but no strong energy dependence is observed. At $\sqrt{s_{NN}}$ = 200 GeV high-$p_{T}$ J/$\psi$ are strongly suppressed in central Au+Au collisions, that may indicate color screening features.
New STAR upgrades will allow to perform more precise open heavy flavor and quarkonia measurements in next years.

\section*{Acknowledgements}

This publication was supported by the European social fund within the framework of realizing the project ,,Support of inter-sectoral mobility and quality enhancement of research teams at Czech Technical University in Prague'', CZ.1.07/2.3.00/30.0034. 

\section*{References}

\bibliographystyle{iopart-num} 
\bibliography{/home/barbara/Work/Bibliography_all}

\end{document}